\pdfoutput=1
\documentclass{aa}

\usepackage{graphicx}
\usepackage{txfonts}
\usepackage{xspace}
\usepackage{ulem}

\def\ie{i.e.\xspace}
\newcommand\Eq[1]{Eq.~(\ref{#1})}
\newcommand\Fig[1]{Figure~\ref{#1}}

\newcommand\Sec[1]{Section~\ref{#1}}
\newcommand\xE{{x_\mathrm{E}}}
\newcommand\yE{{y_\mathrm{E}}}
\newcommand\xiE{{\xi_\mathrm{E}}}
\newcommand\etaE{{\eta_\mathrm{E}}}
\newcommand\uE{u_\mathrm{E}}
\newcommand\vE{v_\mathrm{E}}
\newcommand{\VE}{V_\mathrm{E}}
\newcommand{\thE}{\theta_\mathrm{E}}
\newcommand{\tE}{t_\mathrm{E}}
\renewcommand{\to}{t_0}

\newcommand{\uo}{u_0}

\def\msun{\mathrm{M}_\odot}

\newcommand\piE{\pi_\mathrm{E}}

\newcommand\DS{D_S}
\newcommand\DL{D_L}

\def\VEb{B_{\mathrm{E},1,2,3}}
\def\phiT{\phi_{\mathrm{E,T3}}}
\def\uEA{u_{\mathrm{E}, 1}}
\def\vEA{v_{\mathrm{E}, 1}}
\def\uEB{u_{\mathrm{E}, 2}}
\def\vEB{v_{\mathrm{E}, 2}}
\def\uEC{u_{\mathrm{E}, 3}}
\def\vEC{v_{\mathrm{E}, 3}}
\def\Ft{\Phi_{\mathrm{\mu l}}}
\newcommand\Ftc[1]{\Phi_{\mathrm{\mu l},\ #1}}
\def\Fto{\Phi_{\mathrm{\mu l}_0}}
\def\Fb{\Phi_{\mathrm{B}}}
\def\Fbo{\Phi_{\mathrm{B}_0}}
\def\Fp{\Phi^{(+)}}
\def\Fm{\Phi^{(-)}}

\def\rp{r^{(+)}}
\def\rm{r^{(-)}}
\def\rpm{r^{(\pm)}}

\renewcommand\d{\mathrm{d}}
\newcommand\I{\mathrm{i}}
\newcommand\conj[1]{\bar{#1}} 
\newcommand\lconj[1]{\overline{#1}}

\newcommand\figsimZ[2]{
\begin{figure*}[!t]
\centering
\includegraphics[width= .8\columnwidth]{fig#1_arcs.pdf}
\includegraphics[width= .8\columnwidth]{fig#1_3D.png}\\
\includegraphics[width= .9\columnwidth]{fig#1_2D.png}
\includegraphics[width= .9\columnwidth]{fig#1_phase.png}\\
\vspace{11mm}
\includegraphics[width= .9\columnwidth]{fig#1_diffarcs_2D.png}
\includegraphics[width= .9\columnwidth]{fig#1_diffps_2D.png}
\caption{#2}
\label{fig_fig#1}
\end{figure*}}
\newcommand\figsimZR[2]{
\begin{figure*}[!h]
\centering
\includegraphics[width= .8\columnwidth]{fig#1_arcs.pdf}
\includegraphics[width= .8\columnwidth]{fig#1_3D.png}\\
\includegraphics[width= .9\columnwidth]{fig#1_2D.png}
\includegraphics[width= .9\columnwidth]{fig#1_phase.png}\\
\vspace{11mm}
\includegraphics[width= .9\columnwidth]{fig#1_diffarcs_2D.png}
\hspace{.88\columnwidth}~
\caption{#2}
\label{fig_fig#1}
\end{figure*}}
\begin{document} 

\title{Interferometric visibility of single-lens models: the thin-arcs approximation}

\author{Arnaud Cassan\thanks{arnaud.cassan@iap.fr}}
\institute{Institut d'Astrophysique de Paris, Sorbonne Universit\'e, CNRS, UMR 7095, 98 bis bd Arago, F-75014 Paris, France}

\abstract{
Long  baseline interferometry of microlensing events can resolve the individual images of the source produced by the lens, which combined with the modelling of the microlensing light curve, leads to the exact lens mass and distance. Interferometric observations thus offer a unique opportunity to constrain the mass of exoplanets detected by microlensing, and to precisely measure the mass of distant isolated objects such as stars and brown dwarfs, and of stellar remnants such as white dwarfs, neutron stars, and stellar black holes. Having accurate models and reliable numerical methods is of particular importance as the number of targets is expected to increase significantly in the near future. In this work we discuss the different approaches to calculating the fringe complex visibility for the important case of a single lens. We propose a robust integration scheme to calculate the exact visibility, and introduce a novel approximation, which we call the `thin-arcs approximation', which can be applied over a wide range of lens--source separations. We find that this approximation runs six to ten times faster than the exact calculation, depending of the characteristics of the event and the required accuracy. This approximation provides accurate results for microlensing events of medium to high magnification observed around the peak (\ie a large fraction of potential observational targets).}

\date{Received / Accepted}
\maketitle

\section{Introduction}

        Measuring the mass of isolated objects in our Milky Way is a major challenge in astrophysics, whether it be the mass of stars, brown dwarfs, or stellar remnants such as white dwarfs, neutron stars, or stellar black holes. Few observational techniques allow such measurements to be made with high precision and/or independently of assumptions about the structure of the targeted objects. Gravitational microlensing \citep{Paczynski1986}, based on the deflection of light rays by a lensing body transiting the line of sight of a distant star, provides the solution of choice for measuring the mass of such isolated lenses. The technique allows us to probe objects, intrinsically luminous or not,  up to Galactic scales and independently of the light emitted by the lens itself. 
        
        Microlensing affects the shape and the number of images of the source star, which results in a global enhancement of the total flux received by the observer. As the individual images produced by the microlens cannot be separated by classical telescopes, what is usually measured is the increase (or magnification) in the flux of the source star as a function of time. Nevertheless, when for bright-enough microlensing events, the lensed  images can in principle be resolved with long baseline interferometers \citep{Delplancke2001, DalalLane2003, RattenburyMao2006, CassanRanc2016} since their typical separation is of the order of a milliarcsecond (\ie within the reach of interferometers with baselines of $\sim40$--$100$ metres. A first series of successful interferometric observations was recently made with the Very Large Telescope Interferometer (ESO/VLTI) on microlensing events TCP J05074264+2447555 `Kojima-1' \citep{Dong2019} and Gaia19bld \citep{Cassan2021}. 
        
        To measure the mass of the lens, two quantities must be derived from the observations. The first is  $\piE$, the microlensing parallax; the second is  $\thE$, the angular Einstein ring radius, which is the angular radius of the ring-like image of the source were it to be perfectly aligned with the lens. The mass follows from $M=\thE/\kappa\piE$, where $\kappa=8.144\ {\mathrm{mas}}/\msun$ \citep{Gould2000}. Ground-based observations can access $\piE$ for long-lasting microlensing events, for which the transverse motion of the Earth is significant enough to allow a good parallax measurement, while for shorter microlensing events space-based parallax is in general required to provide a different vantage point from Earth. Classically, $\thE$ can be estimated from the photometric light curve if the spatial extension of source reveals itself by producing noticeable deviations in the light curve and if the source star is well characterised; in the case of bright microlenses, high-resolution adaptive-optics imaging can also access $\thE$ typically 5--10 years after the microlensing event is over, when the background star and the microlens can be resolved individually. As for long baseline interferometry, it provides a direct measurement of $\thE$ by resolving the split images of the source star and measuring their angular separations. An additional constraint on $\piE$ can be  obtained when times series observations are performed as they allow the direction of the relative motion between the lens and the source to be measured \citep{Cassan2021}.

        The modelling of interferometric data requires both robust and efficient numerical methods to compute the microlensing models, with a good control on numerical errors, in order to calculate the wide range of models typically required by Markov chain Monte Carlo (MCMC) algorithms. In this work we discuss in detail the case of  single-lens models. In \Sec{sec_secVis} we propose a new and more efficient approach for the exact calculation of the complex interferometric visibility than   exists in the literature; we also establish a new approximation, called the thin-arcs approximation, which runs six to ten times faster than the exact calculation, and should apply to a large fraction of potential observational targets. In \Sec{sec_applic} we illustrate and discuss the domain of validity of the thin-arcs approximation by comparing it to the exact calculation, and also to the point-source approximation. We   discuss the possible shortcomings of ill-defined parametrisations, and advocate for suitable sets of parameters that depend on the characteristics of the observed microlensing event. Finally, in \Sec{sec_conc} we summarise the main results and discuss the perspectives of optical/infrared long baseline interferometric observations of microlensing events, in particular in the context of recent developments of the  ESO VLTI/GRAVITY instrument.

\section{Visibility of reference single-lens models} \label{sec_secVis}

\subsection{Key concepts and equations} \label{sec_keys}

\begin{figure}[t]
\includegraphics[width=\columnwidth]{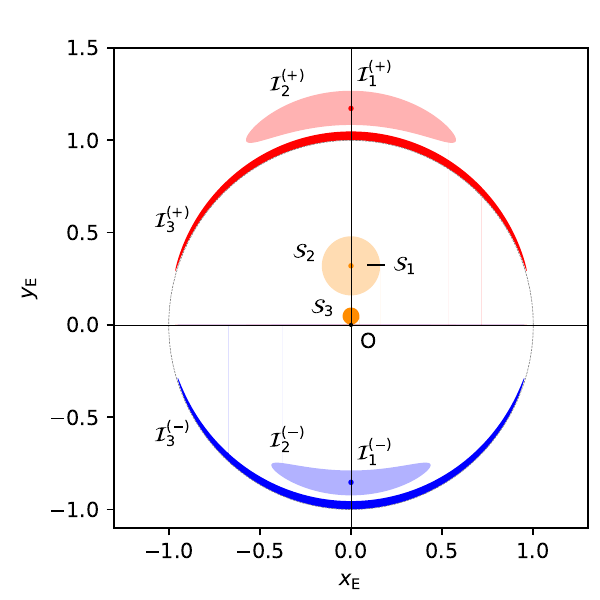}
        \caption{Schematics of the three reference single-lens models detailed in the text. In all cases, the source centre lies on the $\yE$-axis, and the source surface is uniformly bright. Both axes are in $\thE$ units. The microlens is the black dot at the centre of the frame, and the Einstein ring is indicated by the dotted circle of radius unity. In the point-source approximation, the source star $\mathcal{S}_1$ (in orange) is point-like and has two point-like images: the major image $\mathcal{I}^{(+)}_1$ (in red) is located outside of the Einstein ring, while the minor image $\mathcal{I}^{(-)}_1$ (in blue) is   inside. In general, a source   $\mathcal{S}_2$ has a spatial extension, and its major and minor images $\mathcal{I}^{(+)}_2$ and $\mathcal{I}^{(-)}_2$ are elongated along the Einstein ring. When both the source radius and the lens-source separation are smaller than unity, as for source $\mathcal{S}_3$, the two images take the form of two thin arcs, $\mathcal{I}^{(+)}_3$ and $\mathcal{I}^{(-)}_3$. When the microlens lies inside the source, the two images merge into a single ring-like image (not shown here).}
\label{fig_3cas}
\end{figure}

        The lens equation relates the angular position of the background source star to that of its multiple images. If we set up a Cartesian frame of reference $(\mathrm{O}, x, y)$ with axes fixed in the plane of the sky (e.g. north, east) and if we choose the lens to be at the centre of the coordinate system, the complex lens equation for an isolated massive body reads
\begin{equation}
\label{eq_le}
        \zeta = z -\frac{1}{\conj z} \ ,
\end{equation} 
where $\zeta$ is the affix of the (point-like) centre of the source, $z$ the affix of one of the individual point-like images, and $\conj z$ the complex conjugate of $z$. For the lens equation to be correct, the quantities $\zeta$ and $z$ are further normalised by $\thE$, the Einstein angular ring radius \citep{Einstein1936}, which is a function of the lens mass $M$, the observer-lens distance $\DL$, and the observer-source distance $\DS$ through
\begin{equation}
        \thE\equiv\sqrt{\frac{4GM}{c^2} \left(\frac{\DS-\DL}{\DS\DL}\right)} \ ,
\end{equation}
 with $c$ is the speed of light and $G$ the gravitational constant. The typical separation of the images is of the  order of $\thE$; when the source, lens, and observer are perfectly aligned, the image is seen as a perfect ring-shaped image, called an Einstein ring. For a given position $\zeta=u_1e^{\I\theta}$ of the source centre $\mathcal{S}$ (\Fig{fig_3cas}), the single lens equation \Eq{eq_le} is easily solved by writing $z=re^{\I\theta}$, with $r$ solution of $r^2 - u_1r -1 = 0$. This yields two solutions for the images, $z^{(\pm)}=r^{(\pm)}e^{\I\theta}$, where 
\begin{equation}
        r^{(\pm)} = \frac{u_1 \pm \sqrt{u_1^2+4}}{2} \ .
\label{eq_rpmgen}
\end{equation}
If we assume $u_1>0$, the image with $\rp>0$ is the major image ($\mathcal{I}^{(+)}$ in \Fig{fig_3cas}) and that with $\rm<0$ is the minor image ($\mathcal{I}^{(-)}$ in the figure). As $\zeta$, $z^{(+)}$, and $z^{(-)}$ have the same argument, $\mathcal{S}$, $\mathcal{I}^{(+)}$, and $\mathcal{I}^{(-)}$ are aligned together with the lens, as shown in \Fig{fig_3cas}. If the lens is perfectly aligned with the lens (\ie $\zeta=0$), then \Eq{eq_le} yields $|\zeta|=1$ and the image is an Einstein ring of physical angular radius $\thE$. To avoid any confusion in the units in the angular quantities we  discuss here, in the following we assign a subscript `$\mathrm{E}$' to all angular coordinates expressed in $\thE$ units. Hence, we  write $\zeta\equiv\xiE+\I\etaE$ and  $z\equiv \xE+\I\yE$, with 
\begin{equation}
        (\xE, \yE) \equiv (x, y)\ /\ \thE \ ,
\end{equation}
where $(x, y)$ are expressed in radians and $(\xE, \yE)$ in $\thE$ units.

        The complex (fringe) visibility measured by the interferometer is the Fourier transform of the spatial distribution of light $I(\xE, \yE)$ in the plane of the sky (more precisely here, the surface brightness of the images), or
\begin{equation}
        \mathcal{F}[I](\uE, \vE)  \equiv \iint_{\mathbb{R}^2} I(\xE, \yE) e^{-\I 2\pi (\uE \xE+ \vE \yE)} \d\xE\d\yE \ ,
\label{eq_TF}
\end{equation}
where  $(\uE, \vE)$ are the conjugate coordinates of $(\xE, \yE)$. The latter are thus expressed as
\begin{equation}
        (\uE, \vE) \equiv (u, v) \times \thE \ ,
\end{equation}
where $(u, v)$ are the usual spatial frequencies, related to the projected baselines $B_u$ and $B_v$ (respectively in the (O$x$) and (O$y$) directions) by $u=B_u/\lambda$ and $v=B_v/\lambda$, with $\lambda$ the wavelength of observations;   $(u, v)$ are expressed in radians and  $(\uE, \vE)$   in $\thE^{-1}$ units \citep{CassanRanc2016}. Hereafter, we also call `visibility' the quantity  
\begin{equation}
        \VE(\uE, \vE) \equiv \frac{\mathcal{F}[I](\uE, \vE)}{\mathcal{F}[I](0, 0)} \ ,
\label{eq_VEgen}
\end{equation}
which is a normalised version of \Eq{eq_TF}, as the term in the denominator is the total flux. The squared visibility is the squared modulus of the visibility, $|\VE|^2$, and its phase is $\phi=\arg\VE$.

        We now consider a circular uniformly bright source of constant surface brightness $I(\xE, \yE)=I_S$, lensed by an isolated massive body (limb-darkened sources are treated in \Sec{sec_LLD}). As $I_S$ cancels out in the expression of $\VE$ in \Eq{eq_VEgen}, it is convenient to use the quantity $\Phi\equiv\mathcal{F}[I]/I_S$ (which we   also call  `visibility' for simplicity); for the lensed images, we then have
\begin{equation}
        \Ft(\uE, \vE) = \iint_\mathcal{I} e^{-\I 2\pi (\uE \xE+ \vE \yE)} \d\xE\d\yE \ ,
\label{eq_Phi}
\end{equation}
where the subscript `$\mu\mathrm{l}$' stands for microlensed. The integration is performed within the boundaries $\mathcal{I}$ of the (multiple) lensed images. These images are elongated around the Einstein ring, as shown in \Fig{fig_3cas} for the $\mathcal{S}_2$ and $\mathcal{S}_3$ sources.

        In addition to  the lensed images of the source, bright stars in the observing line of sight, although not magnified by the lens, may also be considered as contributing `blend' stars of total visibility $\Fb$. In particular, if the lens itself is bright enough, it may indeed contribute to a blend term $\Phi_L$. Other stars than the lens are unlikely to be involved, even in crowded fields in the Galactic bulge region, because most stars in the immediate vicinity of the lens are faint. We define the blending factor of individual blend star $k$ as the ratio
\begin{equation}
        g_k\equiv F_{B_k}/F_S \ ,
\label{eq_gk}
\end{equation}
where $F_{B_k}$ is the flux of blend star $k$ and $F_S=I_SS$ the flux of the source when it is not lensed, with $S=\pi\rho^2$ and $\rho$ the radius of the source in $\thE$ units. We justify that $I_S$ is used to calculate the flux of both the source and its lensed images, as gravitational lensing has the important property of preserving surface brightness when forming the lensed images. For a given blend star $k$ of surface $S_k$ (in $\thE^2$ units), the star's (constant) surface brightness reads $I_{B_k}\equiv F_{B_k}/S_k=g_kSI_S/S_k$. As in general these stars are not resolved by the interferometer (including the lens;  a typical solar-mass lens at 4 kpc has an angular diameter of about 2 $\mu$as), $S_k$ can be considered infinitely small and we can write $I_{B_k}=g_k S I_S\delta(\xE-\xE_k)\ \delta(\yE-\yE_k)$, with $\delta$ the Dirac distribution and $(\xE_k,\yE_k)$ the coordinates of star $k$. Considering all blend stars, the surface brightness reads
\begin{equation}
        I_B(\xE, \yE) =  \sum_k I_{B_k} = I_S \pi\rho^2 \sum_k  g_k\ \delta(\xE-\xE_k)\ \delta(\yE-\yE_k) \ ,
\label{eq_Ib}
\end{equation}
so that the blend visibility is given by 
\begin{equation}
        \Fb (\uE, \vE) = \pi\rho^2 \sum_k g_k e^{-\I 2\pi (\uE \xE_k+ \vE \yE_k)} \ .
\label{eq_Phib}
\end{equation}
In particular, a bright lens (in the centre of the frame) would contribute to $\Phi_L=g_L\pi\rho^2$, with $g_L=F_L/F_S$ the blend-to-source flux ratio. The overall visibility is expressed as
\begin{equation}
        \VE(\uE, \vE) =\frac{\Ft + \Fb}{\Fto + \Fbo} \ ,
\label{eq_VE}
\end{equation}
where $\Fto\equiv\Ft(0, 0)$ and $\Fbo\equiv\Fb(0, 0)$. When three or more baselines are involved, the bispectrum $\VEb$ and closure phase $\phiT$ of each triangle of baselines are given by 
\begin{eqnarray}
        \VEb &=& \VE(\uEA, \vEA) \times\VE(\uEB, \vEB) \times\VE(\uEC, \vEC) \ ,\label{eq_Bis}\\
        \phiT &=& \arg(\VEb) \ , 
\label{eq_clos}
\end{eqnarray}
where $(\uEC, \vEC) = - (\uEA, \vEA) - (\uEB, \vEB)$ (or an equivalent formula considering that $\VE(-\uE,-\vE)=\lconj\VE(\uE,\vE)$).

        For an interferometer observing in the $H$ band ($\lambda\simeq 1.65\ \mu$m) with projected baseline $B=100$ m along the $x$-axis, and typical values of $\thE$ of $0.5$, $1$, $1.5$, and $2$ mas, the interferometer   probes $\uE$-values of  $0.15$, $0.30$, $0.45$, and $0.60$ (in $\thE^{-1}$ units). As the angular separation of the major and minor images is $\sim2\times \thE$ (see \Fig{fig_3cas}), we expect the visibility to be modulated with a period of $\sim0.5$ along $\uE$, with a first minimum at $\uE\sim0.25$  (in $\thE^{-1}$ units). This means that provided the microlensing event is bright enough to be observed, the lensed images can be resolved in most cases, and the value of $\thE$ measured. 
        
        In the following sections, we derive suitable formulae to compute efficiently \Eq{eq_Phi} for three reference single-lens models: the point-source approximation, the exact formula, and a novel thin-arcs approximation, for uniform and limb-darkened sources.

\subsection{Point-source approximation} \label{sec_PS}
        
        The visibility for a single lens in the point-source approximation was first studied by \cite{Delplancke2001}. The resulting major and minor point-like images can then be modelled by Dirac distributions, respectively located at $\rp$ and $\rm$, from \Eq{eq_rpmgen} and \Fig{fig_3cas}, and weighted by their individual magnification factor $\mu^{(+)}$ and $\mu^{(-)}$ given by      
\begin{equation}
        \mu^{(\pm)} = \frac{1}{2} \left|1\pm \frac{u_1^2+2}{u_1\sqrt{u_1^2+4}}\right| \ .
\label{eq:mupm}
\end{equation}
When there is no source of blend \citep{Delplancke2001}, the squared visibility reads
\begin{eqnarray}
\label{eq:VEsingle}
        |\VE|^2 &=& \left|\frac{\mu^{(+)} e^{-\I 2\pi \vE\rp} + \mu^{(-)} e^{-\I 2\pi \vE\rm}}{\mu^{(+)}+\mu^{(-)}}\right|^2 \nonumber\\
        &=& \frac{1+R^2+2R \cos\left(2\pi\vE\sqrt{u_1^2+4}\right)}{(1+R)^2} \ , 
\end{eqnarray} 
where $R=\mu^{(+)}/\mu^{(-)}$ is the ratio of the magnification of the major image to that of the minor image. The quantity $|\VE|^2$ is a sinusoidal function along the (O$\yE$) direction, and is invariant along the (O$\xE$) direction. The modulation has a period of $T=1/(u_1^2+4)^{1/2}$ and the squared visibility oscillates between $\left(\frac{R-1}{R+1}\right)^2$ and $1$. If $u_1$ is not too large (i.e. $u_1\lessapprox 0.5$), the periodicity is approximately constant and equal to $T\approx0.5$. The amplitude is largest when $u_1$ is small, as $\mu^{(+)}$ and $\mu^{(-)}$ are both approximated by $1/2u_1$ so that $R\approx1$. However, we show in \Sec{sec_applic} that the point-source model provides in this case a poor approximation of the visibility as the lens strongly distorts the images.

        In the general case when one or several unrelated objects $k$ contribute to the blending flux (Sect. \ref{sec_keys}), the resulting complex visibility reads
\begin{equation}
        \VE = \frac{\mu^{(+)} e^{-\I 2\pi \vE\rp} + \mu^{(-)} e^{-\I 2\pi \vE\rm} + \sum_k g_k e^{-\I 2\pi (\uE \xE_k+ \vE \yE_k)}}{\mu^{(+)}+\mu^{(-)} + \sum_k g_k}  \ ,
\end{equation}
where $(\xE_k, \yE_k)$ are the location of the blend sources on the plane on the sky in $\thE$ units, and $g_k$ their blending factors as defined by \Eq{eq_gk}.

        Finally, we note that a single lens event will always lead to a strong interferometric signal, even for microlensing events with relatively low peak magnification or observed far from the peak of the light curve. From the definition of $R$ in \Eq{eq:VEsingle} and \Eq{eq:mupm}, it is possible to compute the maximum amplitude of the oscillation of $\VE^2$, which is equal to $S=1-\left(\frac{R-1}{R+1}\right)^2=4/(u_1^2+2)^2$. Even for an unrealistic observation at very low magnification (e.g. $1.4$; $u_1=0.9$), we have $S>0.5$, and for a more standard value of $u_1<0.2$ (magnification $>5$), the contrast reaches values greater than $S>0.96$.

\subsection{Exact formula for an extended source star} \label{sec_exact}

        The visibility and closure phase for an extended-source single-lens model was first studied by \cite{RattenburyMao2006}. As the images are elongated along the Einstein ring (\Fig{fig_3cas}) with analytically  well-defined contours  (\Sec{sec_keys}), the authors proposed   computing numerically the visibility defined in \Eq{eq_TF} by a line integral along the outer boundary of each of the major and minor images. 
        
\begin{figure}[t]
\includegraphics[width=\hsize]{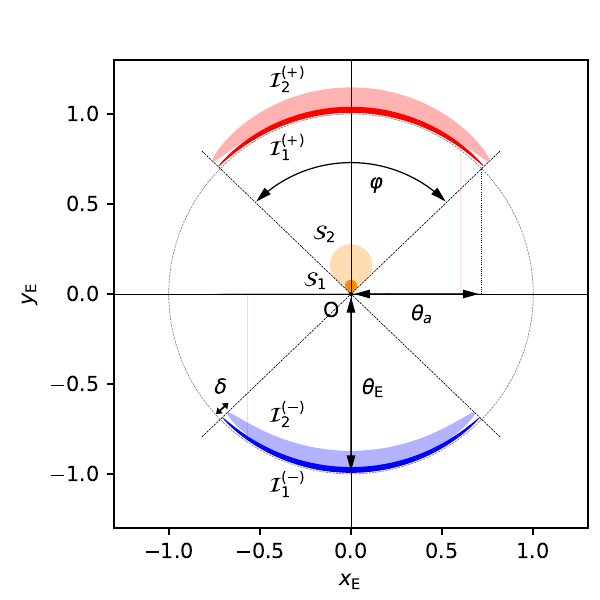}
\caption{Geometry of   major (red) and minor (blue) images when the images are arc-shaped (\ie for $0<\eta_1<1$, where $\eta_1=\rho/u_1$). The Einstein ring is shown as the black dotted circle, and both axes are in $\thE$ units. When the arcs are thin, the images thickness $\delta$ is not resolved by the interferometer and two sources $\mathcal{S}_1$ and $\mathcal{S}_2$ (in orange) of same ratio $\eta_1$ produce arc-shaped images of equal angular elongation, $\varphi = 2\arcsin\eta_1$. The projected extension of the arcs onto the $\xE$-axis is $\theta_a=\eta_1\thE$,  is hence comparable to $\thE$, and is thus resolved by the interferometer. When  $\eta_1>1$, the image of the source is a ring (and $\varphi=\pi$).}
\label{fig_thinarcs}
\end{figure}

        Here, however, we follow a different route and set up another approach to compute \Eq{eq_TF}, motivated by different arguments. Firstly, our tests have shown that while the line integration scheme works well when the images are reasonably elongated (\ie at low to medium magnifications), when the images take the form of thin arcs the parametrisation of the contours is far from  optimal, as the individual points defining the images contours become strongly unevenly spaced. To achieve  reasonable accuracy,   the points on the contours need to be resampled, and in any case the number of contour points must be significantly increased. This operation is mandatory, as the line integral basically operates a subtraction between the wedge-shaped area subtended by the outer boundary and that subtended by the inner boundary, which are both of the order of $\varphi/2$ (as they are located at $r\simeq1$), where  $\varphi$ is the opening angle defined in \Fig{fig_thinarcs}. In contrast, when $\rho\ll 1$ the area enclosed in a given image is of the order of $\rho\varphi$, and the fractional difference corresponding to the searched visibility is of the  order of $\rho\varphi/(\varphi/2)=2\rho\ll1$. Hence, an accurate visibility requires     a very high accuracy on the line integrals, in particular in the portions where the contours are not well sampled by the parametrisation. The approach we derive below is both robust and computationally much more efficient, and allows a precise control on the final accuracy. This can be achieved by calculating \Eq{eq_TF} as a two-dimensional integral in polar coordinates, as we detail below.

\begin{figure*}[t]
        \includegraphics[width=\columnwidth]{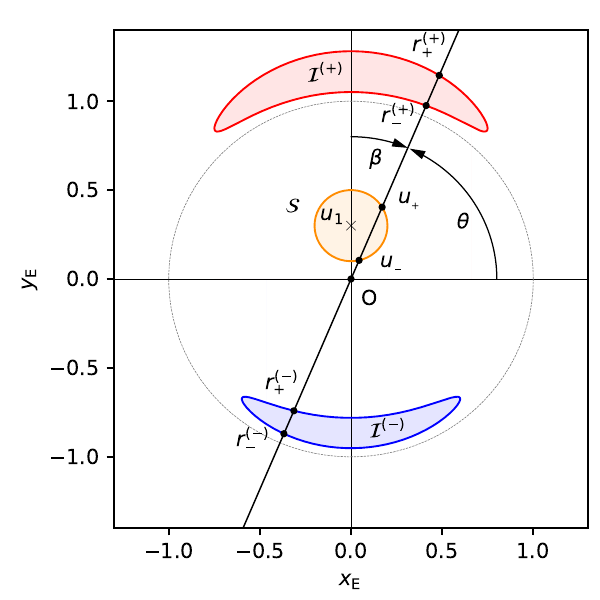}
        \includegraphics[width=\columnwidth]{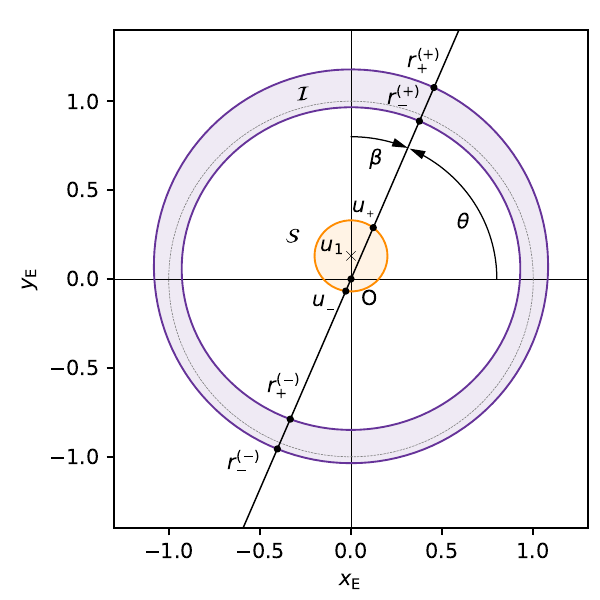}
        \caption{Geometry and parameters used to  calculate   interferometric visibility of an extended source star lensed by a single lens. In both panels the Einstein ring is shown as  the black dotted circle, and both axes are in $\thE$ units. $u_1\geq0$ is the position of the centre of the source along the $\yE$-axis, $u_+$ and $u_-$ are the intersections of the black line (polar angle $\theta$) with the contour of the source, and $(\rp_+, \rm_+)$ and $(\rp_-, \rm_-)$ are their respective images given by the single-lens equation. Arc-like images are shown in the left panel ($0<\eta_1<1$, where $\eta_1=\rho/u_1$), while the case of a single ring-like image ($\eta_1>1$) is shown on the right.}
\label{fig_schema}
\end{figure*}

        We  again let $u_1>0$ be the ordinate of the centre of the source $\mathcal{S}$ along the $\yE$-axis, and $\rho$ the source radius (both in $\thE$ units), as shown in \Fig{fig_schema}. We again assume that the source is uniformly bright (and also the images, since surface brightness is preserved). We let $\theta$ be the angle of the usual polar coordinates, and $u_+$ and $u_-$ the radii where the radial black line in \Fig{fig_schema} intersects the upper and lower contours of the source. From geometrical considerations, these are given by 
\begin{equation}
        u_\pm = u_1\cos\beta\pm\sqrt{\rho^2- u_1^2\sin^2\beta} \ ,
\end{equation}
where $\beta\equiv\theta-\pi/2$. If $\rho<u_1$, we restrict $\beta$ to vary between $-\arcsin\eta_1$ and $\arcsin\eta_1$ (left panel of \Fig{fig_schema}), where
\begin{equation}
         \eta_1\equiv \rho/u_1 \ .
\label{def_eta1}
\end{equation}
Otherwise, if $0<u_1<\rho$, we limit $\beta$ to vary between $-\pi/2$ and $\pi/2$ (right panel). These choices are sufficient to parametrise the two arc-shaped images (or the ring) as we obtain two points above the horizontal axis for the major image $\mathcal{I}^{(+)}$, $\rp_+$, and $\rp_-$, and two points below for the minor image $\mathcal{I}^{(-)}$, $\rm_+$, and $\rm_-$, given by
\begin{equation}
\begin{cases}
        \rp_\pm = \displaystyle\frac{1}{2}\left(u_\pm + \sqrt{u_\pm^2+4}\right) \\[1em]
        \rm_\pm = \displaystyle\frac{1}{2}\left(u_\pm - \sqrt{u_\pm^2+4}\right) \ 
\end{cases}
\label{eq_rpm}
\end{equation}
from \Eq{eq_rpmgen}. In all cases, $\rm_-<\rm_+<0<\rp_-<\rp_+$. If $u_1=0$, the ring is perfectly symmetric (Einstein ring) as $u_\pm=\pm\rho$, so that $\rm_-=-\rp_+$ and $\rp_-=-\rm_+$. To calculate the visibility, we further perform the integration of \Eq{eq_Phi} in polar coordinates, 
\begin{eqnarray}
        \Ft &=& \iint_\mathcal{I} e^{-\I 2\pi r (\uE\cos\theta+ \vE\sin\theta)} r\d r\d\theta \nonumber\\
        &=&  \iint_\mathcal{I} e^{-\I 2\pi \Omega r} r\d r\d\theta \ ,
\label{eq_Phipolar}
\end{eqnarray}
where
\begin{equation}
        \Omega\equiv\uE\cos\theta + \vE\sin\theta \ .
\end{equation}
We also decompose the full integral into two separates integrals, one for the image above the horizontal axis (elongated image or half-ring), $\Fp$, and one for the image below, $\Fm$, with
\begin{equation}
        \Ft = \Fp + \Fm \ .
\end{equation}

        We first consider the case $\rho<u_1$ (or $0<\eta_1<1$). Since $0<\rp_-<\rp_+$ and $\pi/2-\arcsin\eta_1\leq\theta\leq\pi/2+\arcsin\eta_1$, we can write the integral \Eq{eq_Phipolar} for the major image as
\begin{eqnarray}
        \Fp &=& \int_{\pi/2-\arcsin\eta_1}^{\pi/2+\arcsin\eta_1}\left(\int_{\rp_-}^{\rp_+} r e^{-\I 2\pi\Omega r} \d r\right)\d\theta \nonumber\\
        &=& \int_{-\arcsin\eta_1}^{\arcsin\eta_1} R\left(\rp_-, \rp_+\right) \d\beta \ ,
\end{eqnarray}  
where we   change  to variable $\beta$ in the second line,  
\begin{equation}
        \Omega=-\uE\sin\beta + \vE\cos\beta \ ,
\label{eq_Omegabeta}
\end{equation}
and 
\begin{eqnarray}
        R(r_1, r_2) &\equiv& \int_{r_1}^{r_2} r e^{-\I 2 \pi  \Omega r} \d r \nonumber\\
        &=&
        \begin{cases}
        \ \displaystyle\frac{1}{2} \left(r_2^2-r_1^2\right)  & \text{if}\ \Omega=0 \ , \\[1em]
        \ \displaystyle\left[ \frac{e^{-\I 2 \pi  \Omega r} (1+\I 2 \pi  \Omega r)}{4 \pi ^2 \Omega^2} \right]_{r_1}^{r_2} &\text{otherwise}.
        \end{cases}
\label{eq_R}
\end{eqnarray}
The first terms of the series expansion (with respect to $\Omega$) of the expression inside the brackets are 
\begin{equation}
        \frac{e^{-\I 2 \pi  \Omega r} (1+\I 2 \pi  \Omega r)}{4 \pi ^2 \Omega^2} \approx \frac{r^2}{2}+\frac{1}{4 \pi ^2 \Omega ^2} - \I \frac{2}{3}  \pi  r^3 \Omega - \frac{1}{2} \pi ^2 r^4 \Omega ^2 \ .
\label{eq:DL}
\end{equation}
When $\Omega\ll1$, the term $1/4 \pi ^2 \Omega^2$ becomes large, possibly generating numerical issues, although theoretically  this term cancels out in the difference \Eq{eq_R}. Hence, the formula $(r_2^2-r_1^2)/2$ may be used for values of $\Omega$ below a threshold of typically $\Omega\sim10^{-3}$ if we want the second term of the series (in $\Delta r^3$) to contribute no more than $\sim10^{-3}$ times the term in $\Delta r^2$. 

        We proceed in a similar way for the minor image, but this time with $\rm_-<\rm_+<0$ and $-\pi/2-\arcsin\eta_1\leq\theta\leq-\pi/2+\arcsin\eta_1$, so that the integral reads
\begin{eqnarray}
        \Fm &=& \int_{-\pi/2-\arcsin\eta_1}^{-\pi/2+\arcsin\eta_1}\left(\int_{-\rm_+}^{-\rm_-} r e^{-\I 2\pi\Omega r} \d r\right)\d\theta \nonumber\\
        &=& \int_{-\arcsin\eta_1}^{\arcsin\eta_1}\left(\int_{-\rm_+}^{-\rm_-} r e^{\I 2\pi\Omega^\prime r} \d r\right)\d\beta^\prime \ ,
\end{eqnarray}
after changing the  variable to $\beta^\prime=\theta+\pi/2$, and introducing $\Omega^\prime=-\uE\sin\beta^\prime+\vE\cos\beta^\prime=-\left[\uE\cos(\beta^\prime-\pi/2) + \vE\sin(\beta^\prime-\pi/2)\right]=-\Omega$.
As $\beta^\prime$ is a dummy variable, we call it $\beta$ and write 
\begin{eqnarray}
        \Fm &=& \int_{-\arcsin\eta_1}^{\arcsin\eta_1}\left(\int_{-\rm_+}^{-\rm_-} r e^{\I 2 \pi  \Omega r} \d r \right)\d\beta \nonumber\\
        &=& \int_{-\arcsin\eta_1}^{\arcsin\eta_1}\left(\int_{\rm_+}^{\rm_-} r e^{-\I 2 \pi  \Omega r} \d r \right)\d\beta \nonumber\\
        &=& \int_{-\arcsin\eta_1}^{\arcsin\eta_1}\left(-\int_{\rm_-}^{\rm_+} r e^{-\I 2 \pi  \Omega r} \d r \right)\d\beta \ 
\end{eqnarray}
by changing variable $r$ to $-r$ and inverting the boundaries of the integral, so that 
\begin{equation}
        \Fm = \int_{-\arcsin\eta_1}^{\arcsin\eta_1} -R\left(\rm_-, \rm_+\right) \d\beta \ .
\end{equation}

        We now  examine the case $0<u_1<\rho$ (or $\eta_1>1$). In this situation the lens lies inside the source, and there is a single ring-like image. The full ring can be drawn by varying $\theta$ from $0$ to $\pi$, as $\rp_+$ and $\rp_-$ draw the half-ring above the horizontal axis and $\rm_+$ and $\rm_-$ the half ring below it. Hence, the calculation is exactly the same as for $\rho<u_1$;   the only difference is that the integration is now performed between $-\pi/2\leq\beta\leq\pi/2$. 

        In summary, for all values of $u_1>0$ and $\rho>0$ (\ie $\eta_1>0$) we have     
\begin{equation}
        \Ft = \int_{-\beta_m}^{\beta_m} \left[R\left(\rp_-, \rp_+\right) - R\left(\rm_-, \rm_+\right)\right] \d\beta \ ,
\label{eq_Phiex}
\end{equation}
where
\begin{equation}
        \beta_m\equiv\arcsin\left[\mbox{min}\left(\eta_1, 1\right)\right] \ .
\end{equation}
When $u_1=0$ (Einstein ring), this formula still holds with $\beta_m=\pi/2$. In that case the visibility has no imaginary part, which is expected from the symmetry of the ring image.

\subsection{The thin-arcs approximation} \label{sec:flatimages}

        We consider a common case in practice where the source has a small  radius $\rho$  compared to $\thE$ (typically, $\rho\lessapprox0.1$) and passes the lens at small impact parameter, typically $u_1\lessapprox0.2$ (which  corresponds to a point-source magnification at a peak of about $5$). This situation is illustrated in \Fig{fig_thinarcs} for two sources of radii $\mathcal{S}_1$ and $\mathcal{S}_2$. In the figure it is clear that if the major and minor images are resolved by the interferometer along the (O$\yE$) axis (typical angular separation of $2\times\thE$, cf. \Fig{fig_thinarcs}), they have a good chance to be resolved along the (O$\xE$) axis as well (typical angular separation of $2\times\theta_a$, where `$a$' stands for arcs). Since the opening angle of the images is given by $\varphi=2\arcsin\eta_1$ (with $\eta_1=\rho/u_1<1$, and $\varphi=\pi$ for a ring-like image), we have $\theta_a = \eta_1\thE$ (or $\theta_a=\thE$ for a ring). Values of $\eta_1>0.5$ are easily reached for the range of values of $u_1$ we discussed above. Hence, we expect this situation to be common in observed microlensing events.

        While the major and minor images are resolved in their individual elongations ($\sim2\theta_a$) and mutual separation ($\sim2\thE$), on the contrary their thickness ($\delta$ in \Fig{fig_thinarcs}) will most certainly never be resolved by current interferometric facilities: a value of  $\delta$  of the order of  $\rho$  requires  reaching, at best, typical angular resolutions of $\delta\approx\rho\thE<10\ \mu$as. Measuring the thickness of the arcs, however, is not in itself a requirement for measuring $\thE$, as the thickness and the extension of the arcs are directly related by the model. It simply means that the source size $\rho$ is not   measured directly by interferometry, but it does not matter in practice as $\rho$ is usually obtained from the modelling of the photometric light curve. It thus appears natural to investigate the possibility of an approximation formula for the visibility that does not directly depend on $\rho$. A second argument for it is that, as shown in \Fig{fig_thinarcs}, two sources of different radii but with same opening angle $\varphi$ (\ie the same value of $\eta_1$) are difficult to distinguish from an interferometric point of view, as the displacement is again of the order of $\rho\thE$. Hence, a natural model parameter for the  sought-after approximation is $\eta_1$ (\ie the ratio of $\rho/u_1$ instead of the parameters $\rho$ and $u_1$ individually). Conversely, if  $\rho$ to $u_1$ are used in the situation of arc-shaped images (using the exact formula given in  \Sec{sec_exact}), we   expect these parameters to be strongly correlated (if not degenerate), which may alter the smooth running of the fitting process when modelling interferometric data. 
        
\begin{figure}[t]
        \includegraphics[width=\columnwidth]{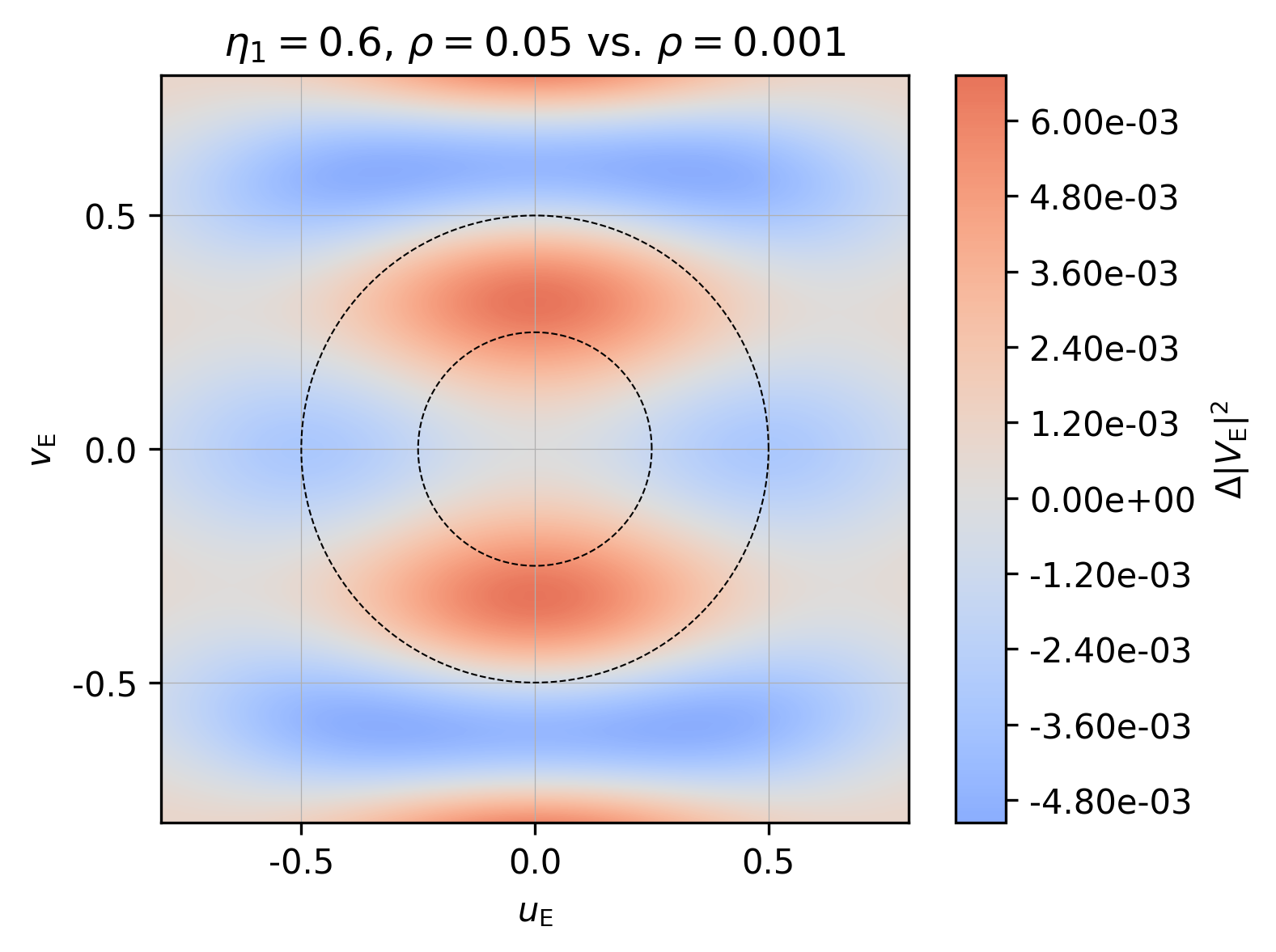}
        \caption{Difference in squared visibility between two single-lens models with same parameter $\eta_1=0.6$. Both axes are in $\thE^{-1}$ units. The inner dashed circle indicates the typical angular resolution (radius $0.25$) and the outer circle (radius $0.5$) twice the typical resolution. The reference model has a source radius $\rho=10^{-3}$ (with $u_1=1.7\times10^{-3}$), while the second model has a source 50 times larger, $\rho=5\times10^{-2}$ (with $u_1=8.3\times10^{-2}$). In this  case the maximum difference in squared visibility is about $6\times10^{-3}$, in practice well below the noise. It justifies the use of $\eta_1$ in the modelling instead of parameters $\rho$ and $u_1$.}
\label{fig_valid}
\end{figure}

        To further illustrate this aspect, \Fig{fig_valid} shows the difference in squared visibility $\Delta|\VE|^2$ between a model computed for parameters $\rho=0.05$ and $u_1=8.3\times10^{-2}$, and a reference model obtained with $\rho=0.001$ and $u_1=1.7\times10^{-3}$, so that both models have same parameter $\eta_1=0.6$. It appears from the figure that while the source radius is multiplied by a factor of 50 between the two models, the squared visibility is not changed by more than $6\times10^{-3}$ (for a maximum excursion between 0 and 1).
        
        To establish a suitable approximation, which we call the  thin-arcs approximation, we note that when both $\rho$ and $u_1$ are small (see \Sec{sec_examples}), $u_\pm$ are small as well so that we can expand $\rpm_\pm$ to first order in $u_\pm$:
\begin{equation}
\begin{cases}
        \displaystyle\rp_\pm \approx 1 + \frac{u_\pm}{2}  \\[1em]
        \displaystyle\rm_\pm \approx -1 + \frac{u_\pm}{2} \ .
\end{cases}
\label{eq_rexpan}
\end{equation} 
Calculating $R$ in \Eq{eq_R} for $\Omega\geq0$ yields
\begin{equation}
\begin{cases}
        R\left(\rp_-, \rp_+\right) = e^{-\I 2 \pi \Omega} \sqrt{\rho^2- u_1^2\sin^2\beta}  \\[1em]
        R\left(\rm_-, \rm_+\right) =  -e^{\I 2 \pi \Omega} \sqrt{\rho^2- u_1^2\sin^2\beta}  \ ,
\end{cases}
\end{equation} 
which allows us to write, in \Eq{eq_Phiex},
\begin{equation}
        R\left(\rp_-, \rp_+\right) - R\left(\rm_-, \rm_+\right) = 2\cos\left(2 \pi \Omega\right) \sqrt{\rho^2- u_1^2\sin^2\beta} \ .
\end{equation}
From \Eq{eq_Omegabeta}, the real part of $\cos\left(2\pi\Omega\right)$ is the product $\cos(2\pi\uE\sin\beta) \times \cos(2\pi\vE\cos\beta)$, while its imaginary part reads $-\sin(2\pi\uE\sin\beta) \times \sin(2\pi\vE\cos\beta)$. As we integrate from $-\beta_m$ to $\beta_m$, the imaginary part cancels out while the real part is doubled. Finally, for $u_1>0$ the visibility reads\footnote{The denominator of $\VE$ in \Eq{eq_VE}, $\Fto=\Ft(0,0)$, can be written  as $\Fto =  4 \rho E(\beta_m, \eta_1^{-2})$, where $E(\phi, m) = \int_0^{\phi} (1-m\sin^2\theta)^{1/2}\,\mathrm{d}\theta$ is the incomplete elliptic integral of the second kind. The total source flux magnification $A$ being the ratio of the total area of the images $\Fto$ to the area of the source, $\pi\rho^2$, we obtain $A=(4/\pi\rho) E(\beta_m, \eta_1^{-2})$, which is the approximation derived by \cite{Yoo2004}, though with a slightly different definition of $E$ and $z=1/\eta_1$.}
\begin{equation}
        \Ft  =  4 u_1 \int_0^{\beta_m} f(\beta)\ \d\beta \ , 
\end{equation}
where again $\beta_m=\arcsin\left[\mbox{min}\left(\eta_1, 1\right)\right]$ and
\begin{equation}
        f(\beta) = \cos(2\pi\uE\sin\beta)\cos(2\pi\vE\cos\beta) \sqrt{\eta_1^2-\sin^2\beta}  \ ,
\label{eq_Ftgen}
\end{equation}
while for $u_1=0$ (Einstein ring), 
\begin{equation}
        \Ft  =  4 \rho \int_0^{\frac{\pi}{2}}  \cos(2\pi\uE\sin\beta)\cos(2\pi\vE\cos\beta) \ \d\beta  \ .
\label{eq_Ftring}
\end{equation}

        As expected, the integrand \Eq{eq_Ftgen} does not depend on $\rho$ and $u_1$ individually, but on their ratio $\eta_1$ (for the perfect Einstein ring, the integrand \Eq{eq_Ftring} does not depend of any of these parameters). In both cases (assuming no blend stars, or $\Phi_B=0$), the factors $4u_1$ or $4\rho$ cancel out in \Eq{eq_VE}, so that the visibility $\VE$ depends on $\eta_1$ only for $u_1>0$. It is noteworthy that in the thin-arcs approximation, $\phi=\arg\Ft$ can take only two values, $0$ and $\pi$. It means that the bispectrum $\VEb$ in \Eq{eq_Bis} is also real, and the closure phases $\phiT$ in \Eq{eq_clos} are $0$ or $\pi$. As $u_1$ increases, the exact value of $\phiT$ starts to differ from these two values, and should be calculated with the exact formula derived in \Sec{sec_exact}. 
        
        Our numerical simulations show that the thin-arcs approximation speeds up the computation by a factor of $6$ to $10$ (depending on the specific configuration of the images) compared to the exact formula, under a common implementation in Python using the \texttt{scipy/romberg} integration scheme and a given achieved accuracy ($5\times10^{-5}$) on both the real and imaginary parts of $\VE$. As we further discuss in \Sec{sec_examples}, the domain of validity of the thin-arcs approximation is wide. Considering $|\VE|^2$, for the usual values of $\rho$ and typical excursions in the $\uE\vE$-plane, the point-source approximation can be linked with the thin-arcs approximation without having to use the exact formula at all.

\subsection{Stellar limb darkening} \label{sec_LLD}
        
        The most convenient way to treat limb-darkening effects is to decompose the source (assumed to be a disk) into $N$ concentric annuli of inner and outer radii $\rho_{k-1}<\rho_k$ ($1\leq k\leq N$) of constant surface brightness $I_k$, with $\rho_0=0$ and $\rho_N=\rho$. The visibility is then simply calculated as
\begin{equation}
        \Ft = \sum\limits_{k=1}^{N} I_k \left(\Ftc{k}-\Ftc{k-1}\right) \ ,
\end{equation}
where $\Ftc{k}$ is computed for a source of radius $\rho_k$ (or equivalently for the thin-arcs approximation, $\eta_{1,\ k}$, with $\eta_{1,\ 0}=0$ and $\eta_{1,\ N}=\eta_1$). As limb-darkening affects the border of the disk of the source star, it will affect the ends of the arc-shaped images and will contribute as a correction only to the visibility. 

        In any case, a linear limb-darkening law will always provide a suitable description of the source's limb darkening. Adopting the (microlensing) convention that the limb-darkening law is normalised to total unit flux yields $I_k = 1-\Gamma\left(1-\frac{3}{2}\left(1-r_k^2\right)^{1/2}\right)$, where $\Gamma$ is the linear limb-darkening coefficient ($0\leq\Gamma\leq1$), and where $r_k=\rho_k/\rho=\eta_{1,\ k}/\eta_1$. The coefficient $\Gamma$ is related to the more usual coefficient $a$ by $a=3\Gamma/(2 + \Gamma)$. The choice of the particular set of values $\rho_k$ (or $\eta_{1,\ k}$) can be optimised to minimise $N$ (e.g. with a linear sampling of $I_k$ between its maximum and minimum values from centre to limb, respectively $(1+\Gamma/2)$ and $(1-\Gamma))$.

\section{Application} \label{sec_applic}

\subsection{Examples and discussion} \label{sec_examples}
        
        Typical examples of visibilities ($|\VE|^2$ and $\phi=\arg\VE$) are shown in Figs. \ref{fig_fig1} to \ref{fig_fig6}. All the figures were calculated for the same value of $\rho=0.03$, slightly above the typical values to challenge the approximations derived in the previous sections. From \Fig{fig_fig1} to \Fig{fig_fig5} the distance of the source to the lens is decreased from $u_1=0.6$ to $0.032$, and \Fig{fig_fig6} is a perfect Einstein ring ($u_1=0$). For each figure the upper panels show, on the left, the positions and shapes of the source and the images with the lens in the centre and, on the right, a three-dimensional view of the squared visibility in the Einstein $\uE\vE$-plane. The middle panels show the squared visibility $|\VE|^2$ (left plot) and the phase $\phi$ (right plot). The bottom panels show the difference in squared visibility between either the point-source approximation (left) or the thin arcs approximation (right)  and the exact calculation. For reference, the colour scale for these plots is set to saturate at $\Delta|\VE|^2\pm0.1$ (negative values in blue, positive in red) as they are typical minimum instrumental values of error bars on $|\VE|^2$. 
        
        For relatively large source-lens separations, such as $u_1=0.6$ in \Fig{fig_fig1}, the images are just slightly elongated, and the  point source provides a good approximation to the visibility. The minimum squared visibility differs from $0$ and the phase spans a range of values because the two images have different magnifications. The thin-arcs approximation does not  provide a good approximation in that case. When $u_1=0.3$ (\Fig{fig_fig2}), the point-source approximation still holds for typical $(\uE, \vE)$ values probed by the interferometer (data points expected at best between the two dashed circles), and the thin-arcs approximation starts to provide a fair approximation within the inner dashed circle. 
        
        When $u_1\lessapprox0.2$ (peak point-source magnification of about 5), the situation is reversed, as seen in \Fig{fig_fig3} for $u_1=0.1$. The thin-arcs approximation now provides a very good approximation to the squared visibility (error $\leq10^{-2}$) even outside the outer dashed circle. The point-source approximation is no longer a suitable model. In \Fig{fig_fig4} ($u_1=0.05$) and \Fig{fig_fig5} ($u_1=0.032$), the squared visibility progressively takes a circular shape, while the phase does not differ more than a few degrees from $0$ or $180\deg$. Finally, in \Fig{fig_fig6}, the image is a perfect ring ($u_1=0$) and the thin-arcs approximation is as accurate as the exact calculation. The phase takes only the two values $0$ and $180\deg$.
        
        In the situation described here ($\rho=0.03$), the transition from the point-source to the thin-arcs approximation appears smooth for $|\VE|^2$, and unless we  have to model interferometric data with very small error bars, it appears unnecessary to perform the exact calculation. The closure phase, however,   still requires   using the exact calculation, but its value is not expected to deviate more than a few degrees from $0$. Finally, for smaller values of $\rho$, the thin-arcs approximation   gives even better results for $|\VE|^2$, which justifies its use for a wide range of single-lens parameters.

\subsection{Practical modelling} 

\begin{figure}[t]
\includegraphics[width= \columnwidth]{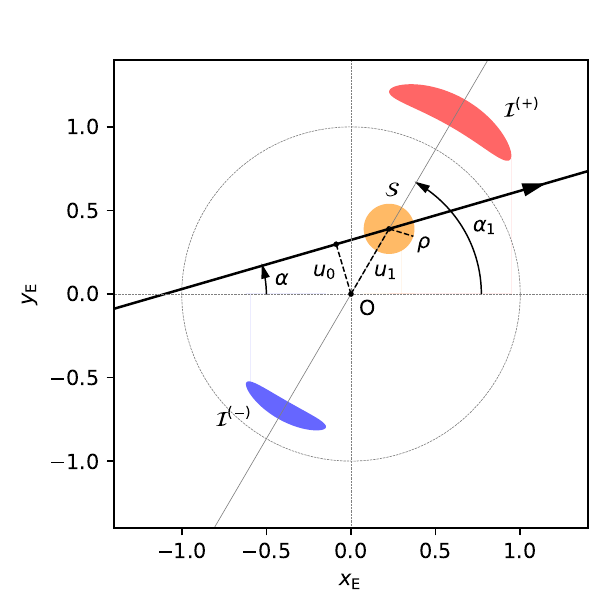}
\caption{Model parameters for a straight-line source-lens trajectory. The lens is the black dot in the centre, the source (unseen) is the orange disk $\mathcal{S}$ of radius $\rho$, and the red and blue arcs $\mathcal{I}^{+}$ and $\mathcal{I}^{-}$ are the major and minor lensed images, respectively. The line joining the centre of the two images makes an angle $\alpha_1$ with the $\xE$-axis, with the convention that the source centre lies at a distance $u_1>0$ from the lens, so that the major image is up for $\alpha_1\in[0, \pi]$ The trajectory of the source with respect to the lens is shown as the  thick black arrow. It makes an angle $\alpha$ with the $\xE$-axis, and $u_0$ is its signed impact parameter. As the source moves relative to the lens along the black arrow, the images rotate around the lens.}
\label{fig_paras}
\end{figure}

        In this section we study possible strategies for fitting interferometric data to single-lens models, and we discuss suitable choices of model parameters for single-epoch or time-series data. In the following, we assume that the limb-darkening coefficient $\Gamma$ of the source can be estimated independently (e.g. from a colour-magnitude diagram). The main parameters we discuss below are shown in \Fig{fig_paras}.
      
        We first consider individual interferometric epochs. We first insist that the images are almost static during the time of an interferometric exposure ($\sim10$ mins). To compute the visibility, the point-source approximation requires $u_1$ as parameter, the exact formula $(u_1, \rho),$ and the thin-arcs approximation $\eta_1$. To fit the interferometric data, in all cases we must add $\alpha_1$ as a parameter (orientation on the sky of the images) as well as $\thE$ (to convert Einstein units into radians). If the lens is luminous, an extra parameter must be added (\Sec{sec_keys}): $g_L$ when using the point-source approximation or the exact formula, and $g_L^\prime=g_L\pi\rho^2/u_1$ when using the thin-arcs approximation. In general, we expect $g_L$ to be at least partly constrained by the light curve, so this parameter may not  necessarily be fitted. 

        In the case of time-series interferometric observations, unless we independently obtain a precise information on the exact trajectory of source relative to the lens, we can assume that the source-lens trajectory is well approximated by a straight line. This approximation holds as long as the different epochs of observation span a relatively short interval of time, typically less than two weeks. The model parameter $\thE$ is again used as it gives the overall angular scale of the problem. When using the exact formula, the source radius $\rho$ must be added as a parameter. Using the lens as a blend star adds another  parameter, $g_L$ or $g_L^\prime$, as described in the previous paragraph. To describe the source-lens straight trajectory, we need to add the parameters $\alpha$, the trajectory angle, and $\to$, a time origin (usually chosen to be the date at which the source is closest to the lens). 
        
        When using the exact formula or the point-source approximation, we also have to add in the list of parameters the Einstein timescale $\tE$ (\ie the time it takes for the source to travel $\thE$) as well as $\uo$, the minimum impact parameter of the source-lens trajectory. However, when using the thin-arcs approximation, $\uo$ is no longer a convenient parameter: if for every individual epoch only $\eta_1=\rho/u_1$  can be measured, then (from Thales' theorem) only $\eta_0\equiv\rho/\uo$ can be used as a parameter of the model. Similarly, $\tE$ cannot be determined individually, as a trajectory with a higher value of $\uo$  implies a slower moving source (\ie a higher value of $\tE$) for (almost) the same shape and location of the images; more precisely, the time $\Delta t$ it takes for the source to travel between two epochs of observation is $\Delta t \propto \uo\tE$, which means the product $p=\uo\tE$ is constant. Hence, in principle $p$ could be used as a model parameter, but in practice it appears more convenient to use the source radius crossing time, $t_*\equiv\rho\tE$, as it is the product of the two constant quantities $\eta_0\times\uo\tE$, and a classical parameter in microlensing modelling.

\section{Summary and perspectives} \label{sec_conc}

        In this work we first reviewed the main concepts and general formulae of interferometric microlensing, and detailed the equations useful for treating the case of a single lens. We recalled the well-known visibility formula for a point source, and then treated the case of an extended source, for which we proposed a new approach for the calculation of the visibility, allowing a robust and numerically efficient calculation. 
        
        This formalism allowed us to establish a new approximation, which we called the thin-arcs approximation, and which applies to microlensing events of medium or higher magnification observed around the peak (\ie a large fraction of potential observational targets). We demonstrated that the computation time using this approximation is six to ten  times faster than with the exact formula, and applies over a wide range of lens-source separations. It even turns out that a direct transition from the point-source to the thin-arcs approximations is possible in many situations, without having to calculate the visibility with the exact formula. 
        
        Accurate models and reliable numerical methods are of particular importance as the number of targets is expected to increase significantly in the near future. Based on a four years of statistics of microlensing events alerted by the OGLE collaboration (2011--2014, about 7\,000 events), \cite{CassanRanc2016} found that the number of potential interferometric targets $N$ scales as $\sim 10^{0.4 \times {\Delta m_K}}$ with the event's peak magnitude $m_K$ \citep[obtained from a linear regression of the event's count, right panel of Figure 3 in][]{CassanRanc2016}; in other words, a gain in $\sim2.4$ magnitudes in the instrument sensitivity results in about ten times more potential microlensing targets. Pushing the limiting magnitude of current or new-generation interferometers will therefore have a huge impact on the field.
        
        Until recently, interferometric facilities  suffered from a lack of sensitivity, limiting the pool of observable microlensing targets to the very bright tip of the distribution. Observations like those obtained for Gaia19bld at the VLTI \citep[peak magnitude of $H=6.2$, just above the PIONIER instrument's limiting magnitude of $H=7.5$, see Ext. Data. Figure 1 in][]{Cassan2021} hence remained exceptional.
        However, the latest improvements in interferometric instruments are going to be a `game-changer', in particular for the ESO GRAVITY instrument at the VLTI. The new dual-field wide mode now available for GRAVITY, which uses a close and brighter star in the vicinity of the target to push the limiting magnitude up to $K\sim16$, will significantly increase the number of potential microlensing targets in upcoming observing campaigns. The expectations are 10 to 30 targets per year (and perhaps more), instead of about 1 target per year with the previous set-up. In addition to pinpointing the masses of exoplanets discovered through microlensing, these observations will allow us, for the first time, to unambiguously find isolated stellar black holes, and measure their masses with an exquisite precision.

\bibliographystyle{aa} 
\bibliography{biblio}



\figsimZ{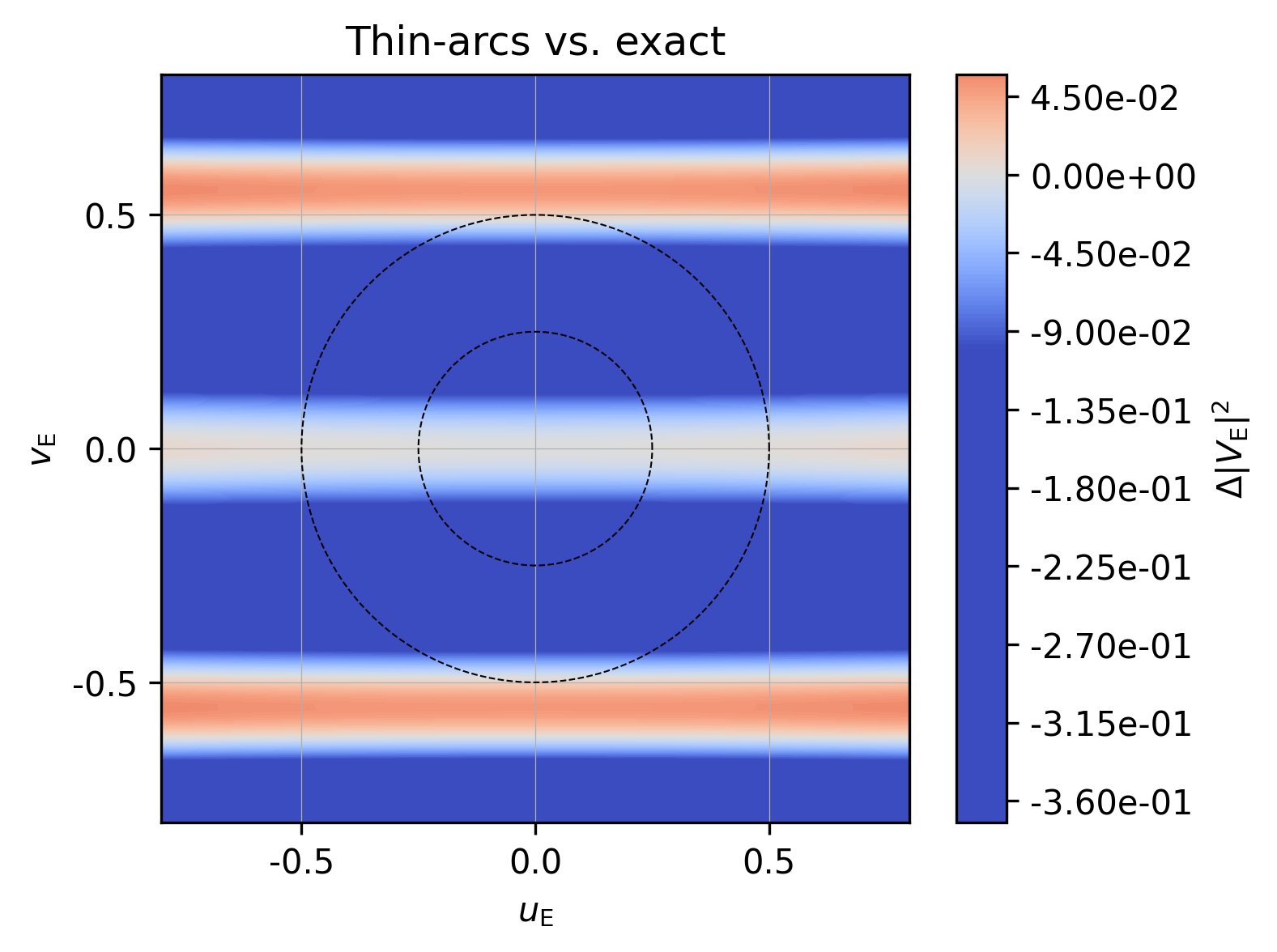}
{Plots for $\rho=0.03$ and $u_1=0.6$.  
\textit{Upper panels:} Shown in the  plot on the left are the lens (black dot in the centre), the (unseen) extended source (orange disk), and its two lensed images (red and blue arc-shaped images), displayed in $(\xE,\yE)$ coordinates which are normalised by the angular Einstein ring radius $\thE$. The plot on the right displays a three-dimensional view of the squared visibility $|\VE|^2$ in the Einstein $\uE\vE$-plane, normalised by $\thE^{-1}$. 
\textit{Middle panels:} Shown in the plot on the left is a contour plot of $|\VE|^2$, while the plot on the right shows the phase of the complex visibility $\phi=\arg\VE$ (the thin white line is a visualisation artefact when $\phi$ jumps from $-\pi$ to $\pi$, or vice versa). 
\textit{Lower panels:} Difference in squared visibility $\Delta|\VE|^2$ between either the point-source (left) or the thin-arcs (right) approximation and the exact calculation. The colours saturate for a difference of $\pm0.1$. The inner dashed circle marks the typical angular resolution (radius $0.25$), and the outer circle (radius $0.5$) twice the typical resolution.} 

\figsimZ{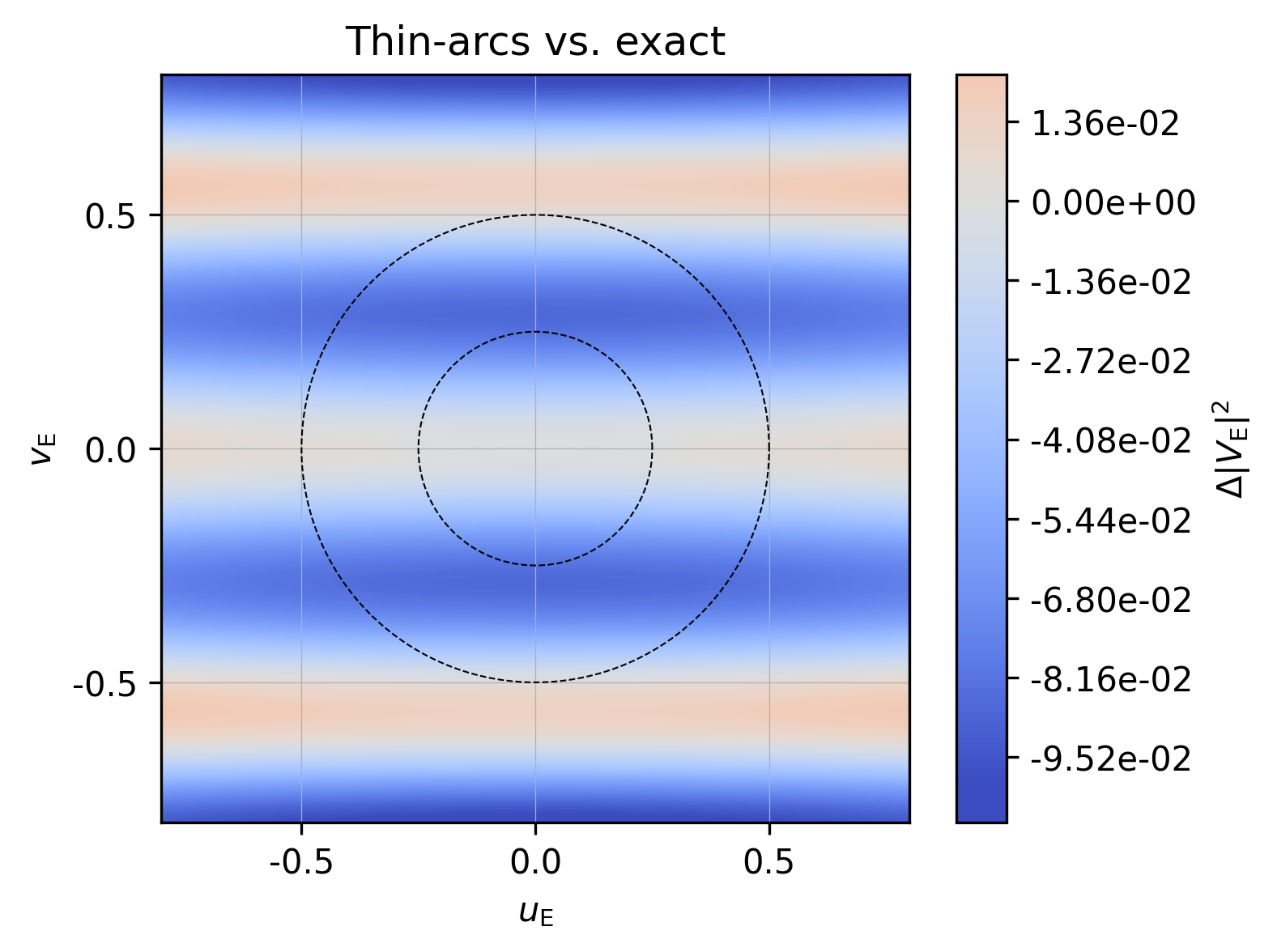}{Same as \Fig{fig_fig1}, but for $u_1=0.3$.}

\figsimZ{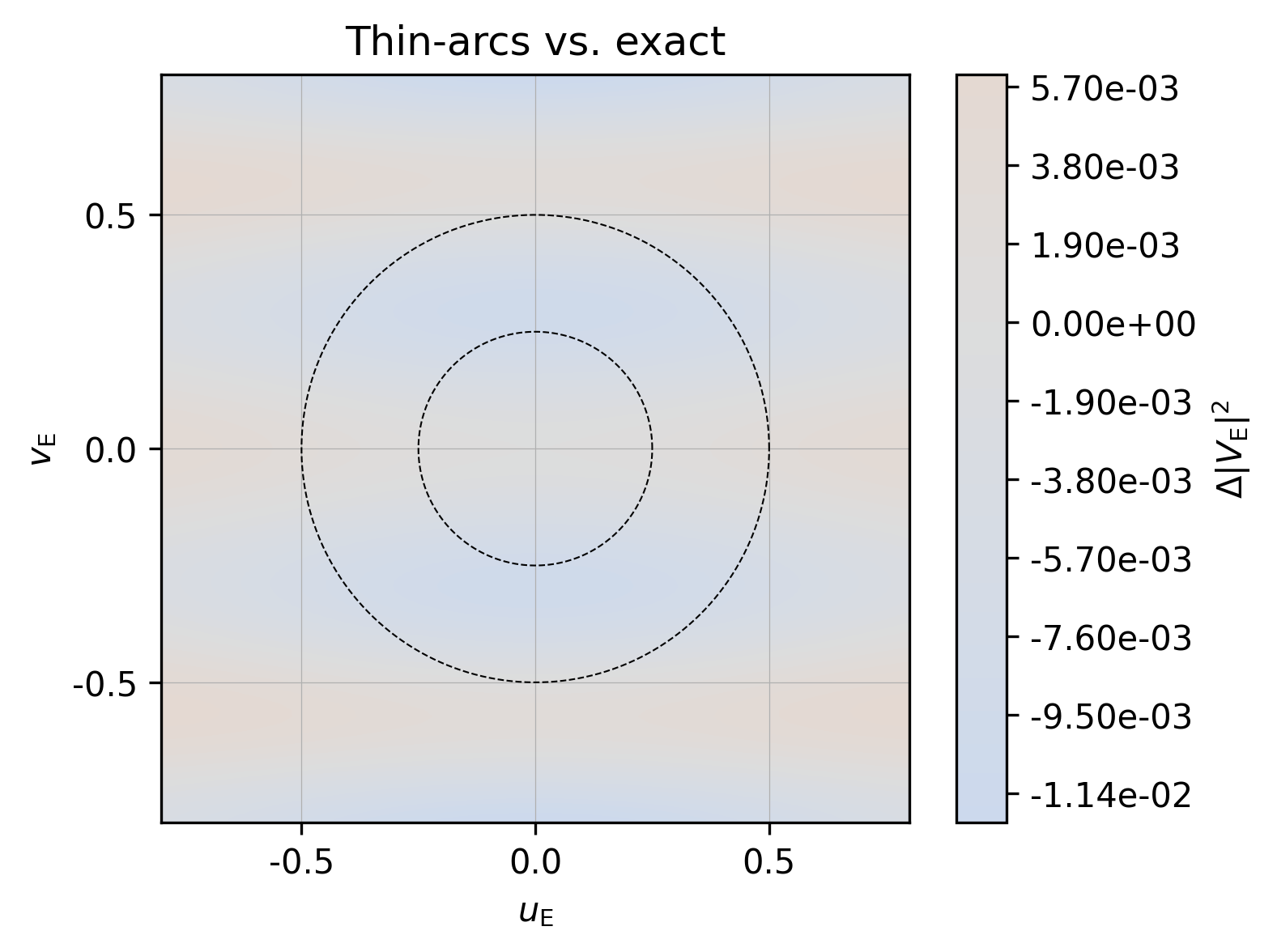}{Same as \Fig{fig_fig1}, but for $u_1=0.1$.}

\figsimZ{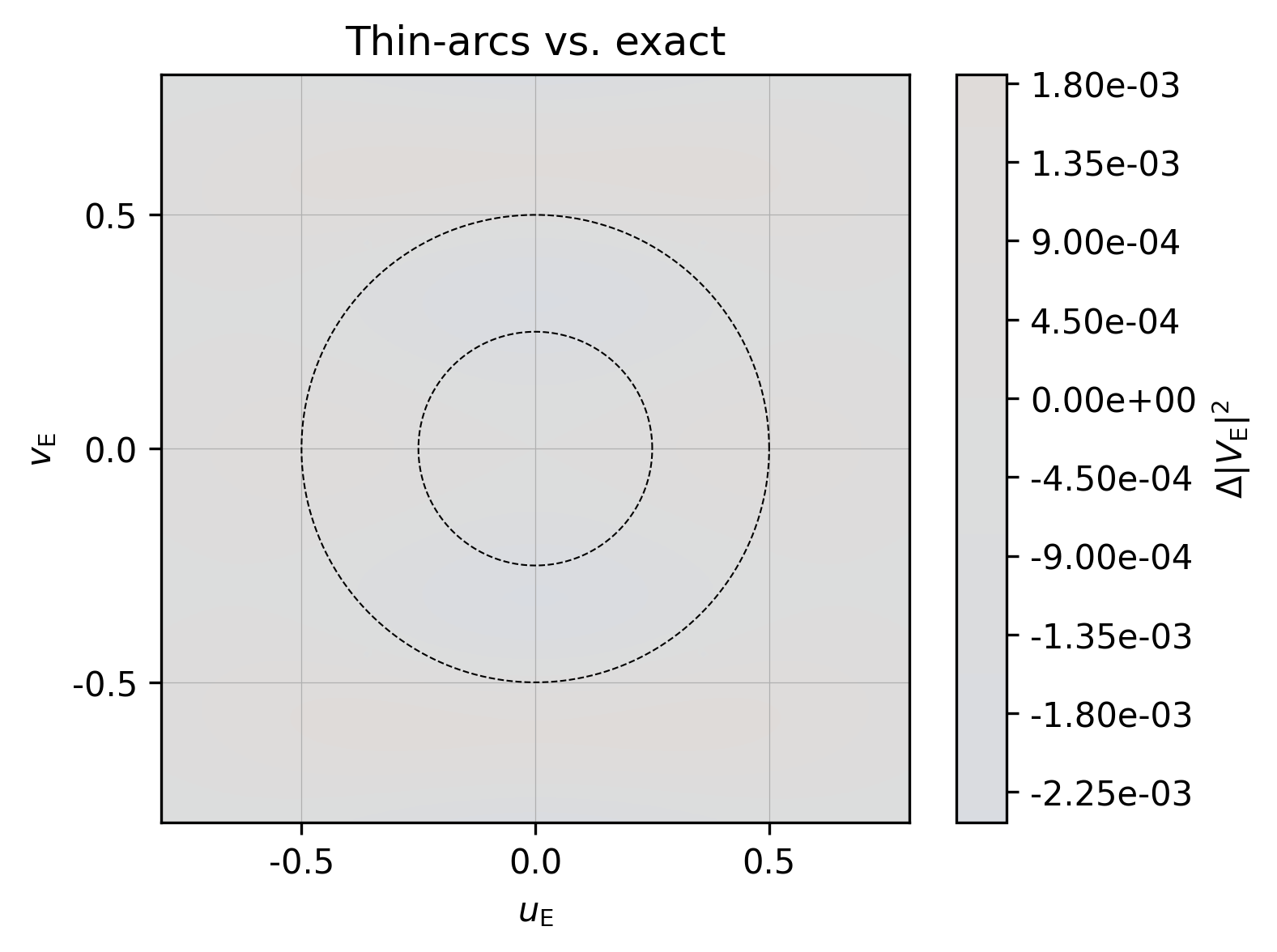}{Same as \Fig{fig_fig1}, but for $u_1=0.05$.}

\figsimZ{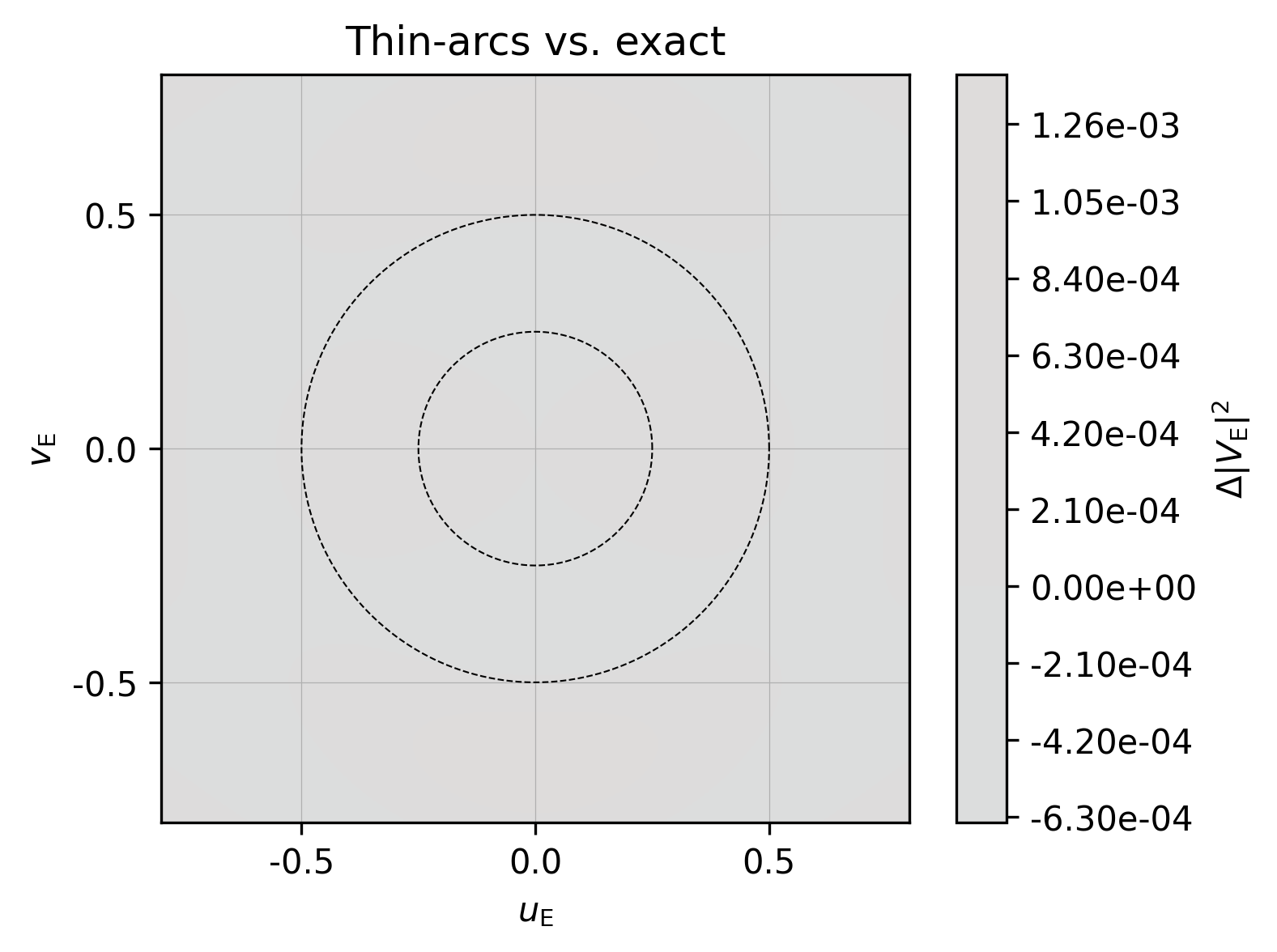}{Same as \Fig{fig_fig1}, but for $u_1=0.032$.}

\figsimZR{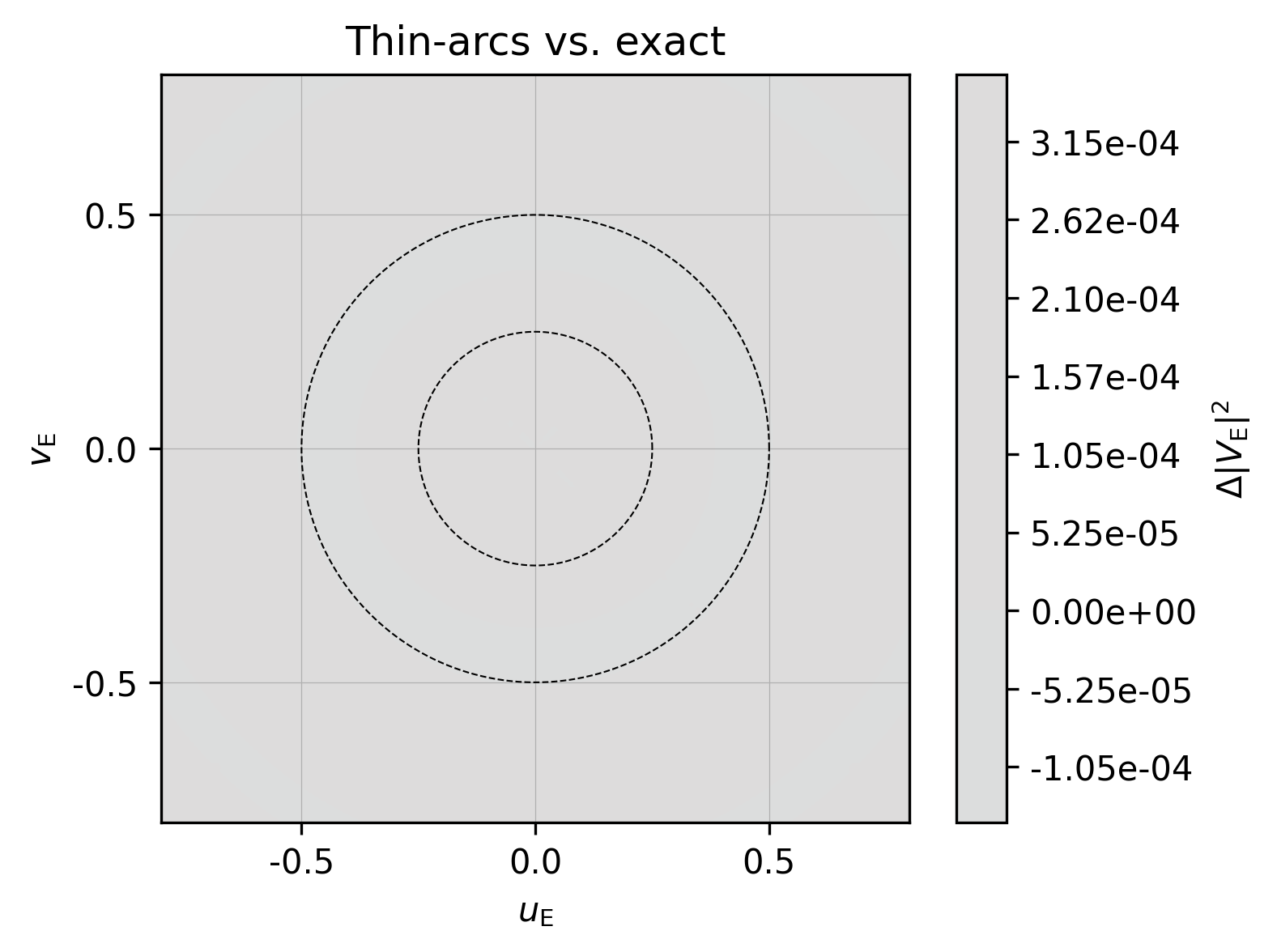}{Same as \Fig{fig_fig1}, but for $u_1=0$. In this case the point-source approximation yields  a circle instead of two point-like images and is not shown here.}


\end{document}